\documentclass[11pt,twoside]{article}
\usepackage{Perlman,epsf}
\def\edcomment#1{\iffalse\marginpar{\raggedright\sl#1\/}\else\relax\fi}
\reversemarginpar

\begin{document}
\title{Optical Structure and Physics of the M87 Jet}
\author{Eric S. Perlman}
\affil{Joint Center for Astrophysics, University of Maryland, Baltimore County,
1000 Hilltop Circle, Baltimore, MD  21250  USA}
\author{William B. Sparks, John Biretta, Duccio Macchetto}
\affil{Space Telescope Science Institute, 3700 San Martin Drive, Baltimore, MD 
21218  USA}
\author{J. Patrick Leahy}
\affil{University of Manchester, Jodrell Bank Observatory, Macclesfield,
Cheshire, SK11 9DL, UK}

\begin{abstract}

We summarize HST observations of the M87 jet, concentrating on polarimetry and
spectral index maps, and compare its optical and radio structures.  The
evidence now supports a stratified model for the structure of the jet, whereby
high-energy, optical synchrotron emitting particles occupy physically different
regions of the jet, closer to the jet axis, with different magnetic field
configurations.  It is in these regions where the shocks that produce the knots
in the inner jet appear to originate. Knot regions have optical spectra which
are much flatter than average for the jet, with the flattest-spectrum regions
coinciding with flux maxima of knots.  These same regions are preceded by
regions where perpendicular magnetic fields are seen. Thus not only do we see
all the necessary ingredients for  {\it in situ} particle acceleration in the
knots, but there is now fairly direct evidence for it as well.  By tracking the
changes in radio-optical and optical spectral index in the knot regions, we can
comment on  acceleration and cooling timescales in each knot.  

\end{abstract}

\section{Introduction}

The giant elliptical galaxy M87 hosts the best-known extragalactic
jet.  Because of its proximity (D=16 Mpc) and high surface brightness, in M87
we can study jet physics in far higher detail than is possible
for any other source.  

Since its launch, HST has studied M87 in great detail.  A comparison of the
jet's
radio and optical structure (Sparks et al. 1996) finds
remarkable large-scale similarity, but clear and significant differences appear
on $0.1''$ scales.  The optical emission is more concentrated in knots and
along the centerline, a trend which continues into the UV.  All of the
inner jet knots (i.e., interior to knot A), show apparent superluminal motion,
with speeds ranging from $2-6 c$ (Biretta et al. 1999).  The superluminal
components vary significantly on 1 year timescales.

With this in mind, we discuss polarimetric and spectral index maps of the M87
jet.   Polarimetry is by far the most sensitive probe of a jet's structure,
because it tracks directly the structure of the magnetic field in the emission
regions.  The comparison of polarized structures in radio and optical bands is
a particularly powerful tool, as it allows us to see whether the magnetic
fields in those regions have similar or different properties.    Spectral index
images are  equally powerful, as they allow us to view not only the spectrum of
the emitted radiation, but also the  underlying spectrum of the synchtron
emitting particles.   A  comparison of spectral index maps in two bands can
diagnose the structure of the jet and the presence of particle  acceleration. 

Here we emphasize certain features in the polarization and
spectral index images presented in Perlman et al. (1999, 2001), and discuss
in particular the inner jet.  For more detailed discussions, the reader is
directed to those papers.

\section{Maps and Results}

In Figure 1, we show a comparison of radio and optical polarimetry in the inner
regions of the M87 jet.  Figure 2 compares the run of radio-optical and optical
spectra with that of optical flux in these same regions.  

\begin{figure}
\plottwo{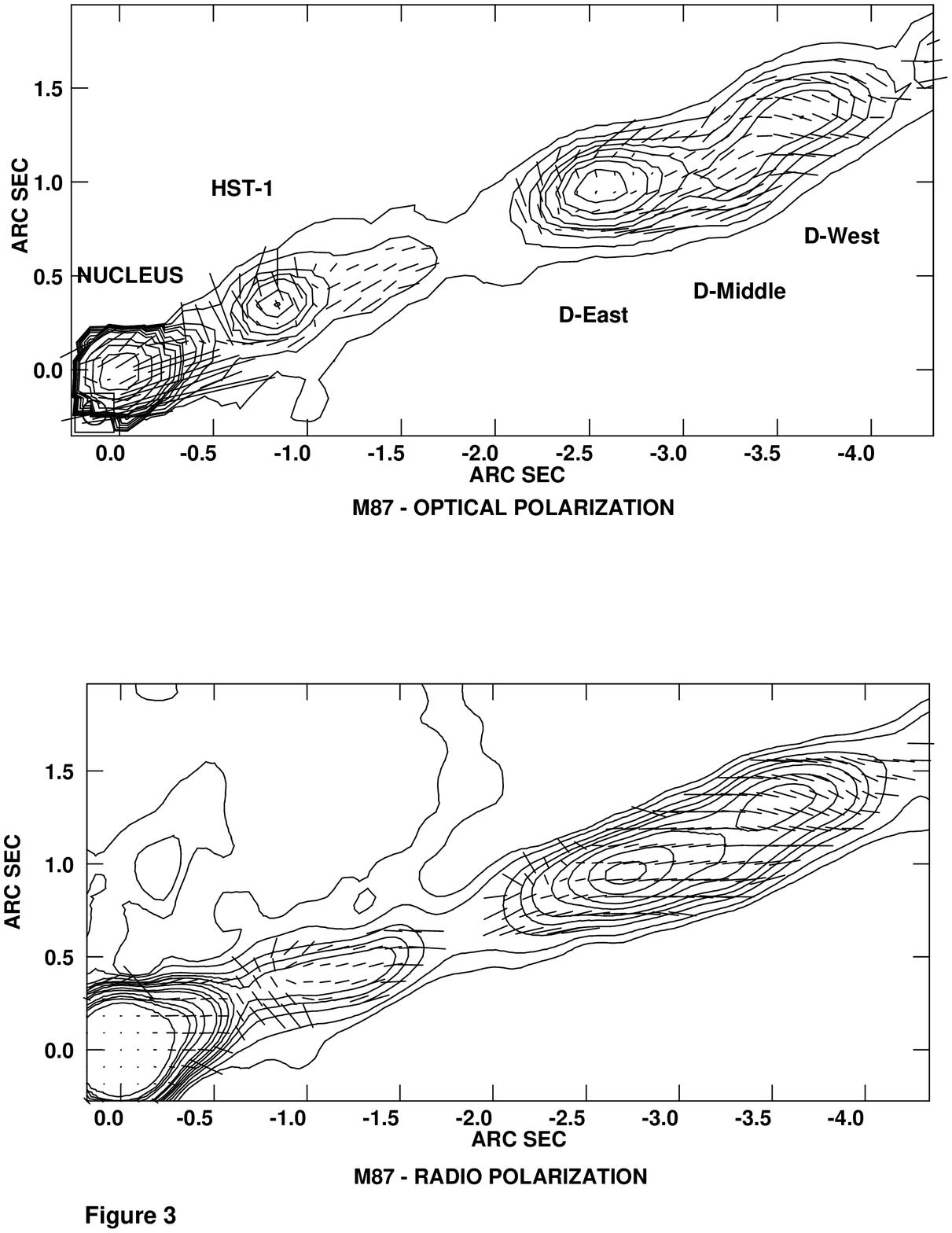}{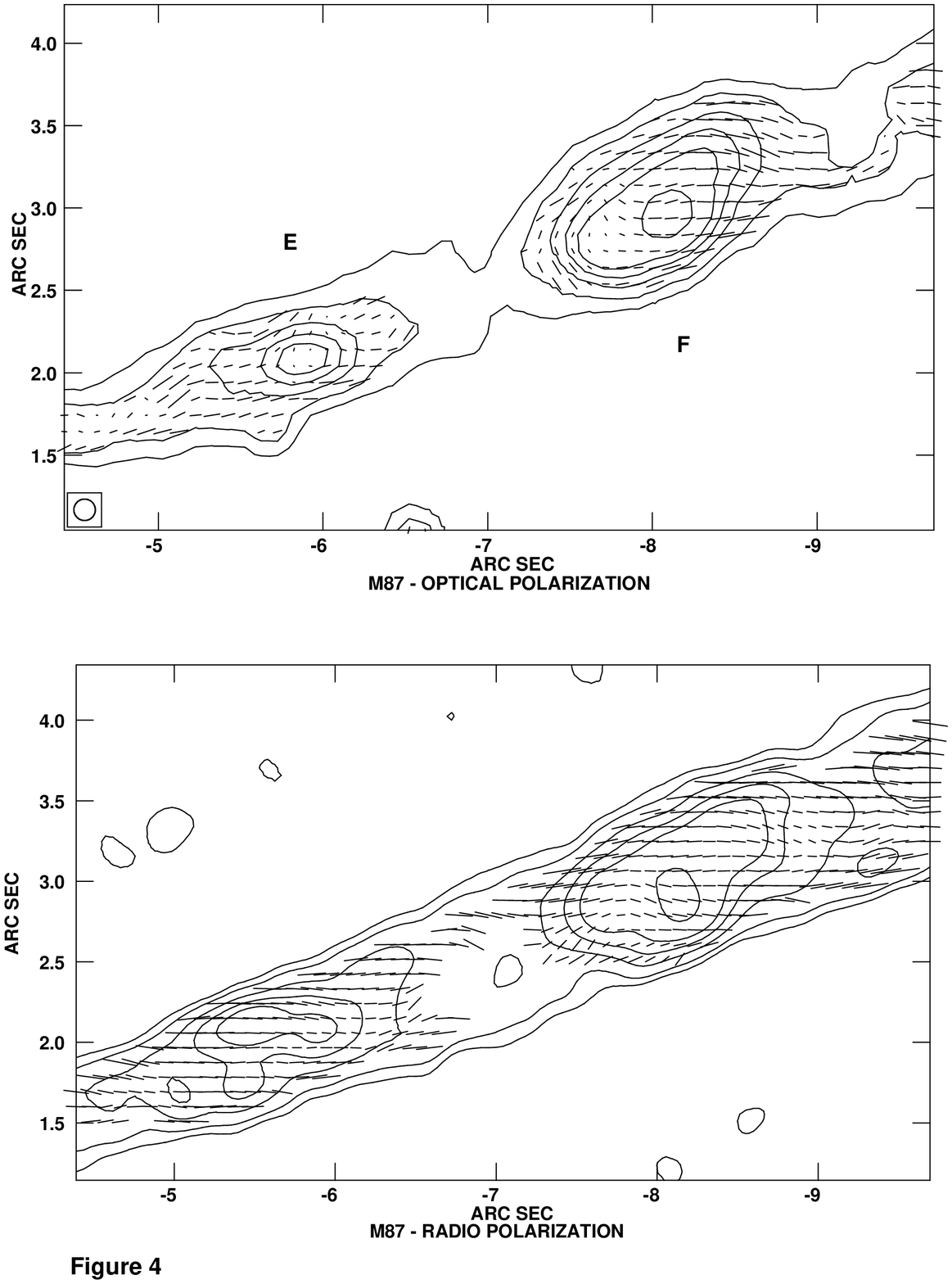}

\caption[]{Polarization maps of the inner jet.   North is at the top and
East is at the left.  The optical image is contoured at (1, 2, 4, 6, 8, 12, 16,
24, 32, 64, 128, 256) $\times 50$ ADU/pix, while the radio image is contoured
at (1, 2, 4, 6 ,8, 12, 16, 24, 32, 64, 128, 256) $\times 0.5$ mJy/beam.  The 
vectors represent MFPA ($1''$ = 300\% polarization).}
\end{figure}

Significant differences are evident in the optical and radio polarization
structures of  knots in the inner jet (Figure 1).  The polarization varies far
more in the optical than it does in the radio, with the strongest variations
near the flux maxima. The average optical polarization just upstream of the
flux peak in HST-1 is $<P_{opt}> = 0.45 \pm 0.08$, while at the maximum,
$P_{opt}$ drops to just $0.14 \pm 0.05$.  In D, E and F, $P_{opt}$ values of
$0.15-0.30$ are observed just upstream of the flux peaks, compared with $<0.10$
at the flux peaks. By comparison, $P_{rad}$  drops from $\sim 0.25$ to 0.15 at
the flux peak, a similarly small change is observed in F, and the  drops in
$P_{rad}$ observed at the flux peaks of D and E are statistically
insignificant.  Obvious differences in the magnetic field orientation are also
observed. Notably, at the upstream ends of knots HST-1, D and F, the optical
magnetic field position angle (MFPA) becomes nearly perpendicular to the jet,
whereas the radio MFPA remains primarily along the jet axis in D and F.

\begin{figure}
\plotone{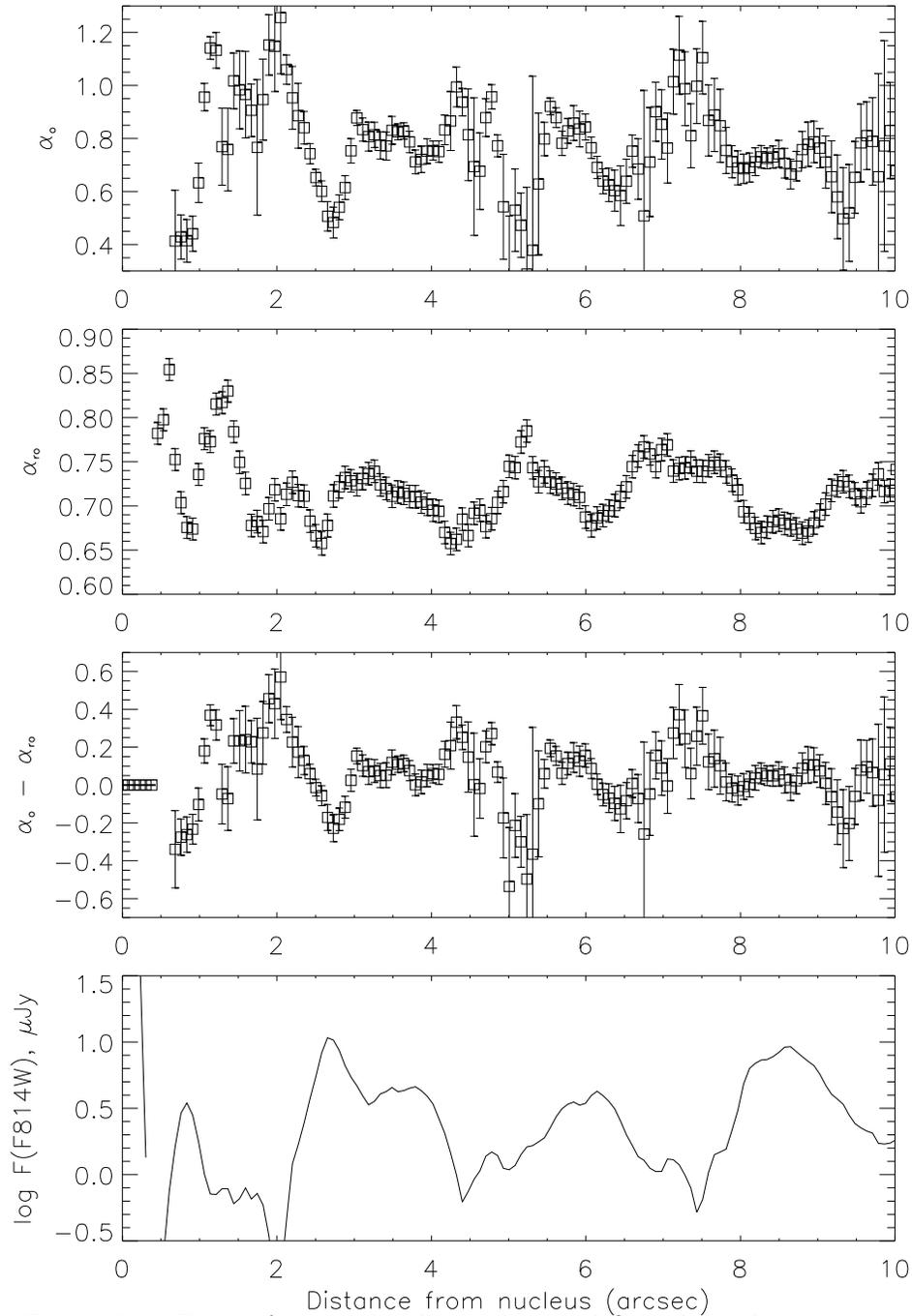}

\vskip 0.15in

\caption[]{Runs of spectral index and optical flux along the jet axis.   
Only the inner jet region is shown.  As can be seen, 
much stronger spectral variations are seen in $\alpha_o$ than$\alpha_{ro}$, and
the two spectral indices do not vary in a correlated fashion near flux maxima.}
\end{figure}

We also see clear differences in these regions between variations in $\alpha_o$
and $\alpha_{ro}$ (Figure 2).   Much larger variations are observed in
$\alpha_o$ near flux maxima, than are seen in $\alpha_{ro}$.  Bright regions of
the knots have flatter $\alpha_o$, with particularly sharp variations in HST-1
and D. Those same regions have much smoother variations in $\alpha_{ro}$. 
Also,  $\alpha_o$ and $\alpha_{ro}$  do not vary together in the knot regions. 
In two regions,  $\alpha_o$ appears to lead $\alpha_{ro}$:  knot HST-1 (by
$0.15''$) and  knot F (by $0.2''$).  But two other knots show just the
opposite, with  $\alpha_o$ lagging $\alpha_{ro}$:  knot D (by $0.15''$) and
knot E (by $0.25''$) In addition, the flat-$\alpha_{ro}$ regions are larger
than the corresponding flat-$\alpha_o$ regions, and significantly larger values
of $\alpha_{ro}-\alpha_o$ are observed immediately downstream of the flux
maxima.  But further downstream this steepening stops once the local mean in
$\alpha_{ro}-\alpha_o$ is reached.

\section{Discussion}

These data are hard to explain unless the jet is stratified, with high-energy
electrons concentrated closer to the axis than lower-energy electrons (Perlman
et al. 1999).  Shocks appear to originate in the high-energy emitting
regions, and thus we see rotations of the optical MFPA just upstream of the
flux maximum.  But at knot maxima,  we see enhanced emission from both the jet
center and edge, such that the polarization vectors cancel out in the optical
but not in the radio.  This model also predicts the much narrower
flat-$\alpha_o$ regions that we see, and the relatively poor correlation
between $\alpha_o$ and $\alpha_{ro}$.

Given the perpendicular MFPAs we observe, the knots posssess  all the necessary
ingredients for localized particle acceleration.  Evidence for particle
acceleration can be seen in the run of $\alpha_o$ (Figure 2), which shows sharp
flattenings beginning in the same regions where  perpendicular MFPAs are
observed, and the flattest spectra coinciding with flux maxima.  It is hard to
explain the runs of $\alpha_o$ and $\alpha_{ro}$ without invoking particle
acceleration, because otherwise one would expect (1) no sharp variations in
$\alpha_o$ near flux maxima, (2)  $\alpha_o$ and $\alpha_{ro}$ should vary
together, and (3)  $\alpha_{ro}-\alpha_o >0$ everywhere in the jet.  Particle
acceleration can easily account for  the long-standing discrepancy
(Meisenheimer et al. 1996 and refs. therein) between the particle ages ($\sim
100-300$ years) and length of the jet (2 kpc), without requiring that the jet
be as strongly out of equipartition as suggested by Heinz \& Begelman (1997). 
Indeed, particle acceleration is expected in these knots if they are analogous
to flaring regions in blazar jets.

If particle acceleration is present, the runs of $\alpha_o$ and $\alpha_{ro}$
represent a time history of particle acceleration and synchrotron aging.   In
two knots (D-East and E), we observe $\alpha_{ro}$ leading $\alpha_o$, while in
two others (HST-1 and F), the opposite  is observed.  Both  are consistent with
models for flares produced by shock-induced particle acceleration.  As Kirk et
al. (1998) showed, the evolution of spectral index in a given band during a
flare depends on the relationship of the acceleration and cooling timescales
$t_{acc}$ and $t_{cool}$ for electrons which emit in that band.  Where
$t_{acc}$ is much shorter than $t_{cool}$, higher-energy emissions should lead
lower-energy emissions (Georganopoulous \& Marscher 1998, Takahashi et al.
1999).  But if $t_{acc}$ and $t_{cool}$ are close to equal near the peak in
$\nu F_\nu$, things are more complex.  At low energies, where $t_{acc}$ is much
faster than $t_{cool}$ (n.b.,$t_{cool} \propto E_\gamma^{-1/2}$), the spectrum
should flatten in advance of the flux maximum, peaking slightly before flux
maximum, followed by a steepening to values somewhat below the jet's nominal
value after maximum flux, and then a return to the nominal value after the
disturbance has passed.  But at high energies, where $t_{cool} \sim t_{acc}$,
spectral {\it steepening} is predicted in advance of the flux maximum, followed
by a hardening as the number of newly accelerated electrons build up.  Under
this model (assuming no injection of fresh particles), flares represent the
reacceleration of old particles within the jet, so they will propagate from low
to high energies, {\it i.e.} observed spectral changes at higher energies will
lag those at lower energies.  We appear to be observing {\it both} situations
in different knot regions in the M87 jet.

\end{document}